\documentclass[submission,copyright,creativecommons]{eptcs}
\providecommand{\event}{AFL 2017}
\usepackage{graphicx}
\usepackage{amsmath}
\usepackage{amssymb}
\usepackage{mathtools}
\usepackage{subcaption}
\usepackage[usenames]{color}

\newcommand{\match}{\mathsf{match}}
\newcommand{\fail}{\mathsf{fail}}

\newcommand{\Rule}{\mathcal{R}}
\newcommand{\NonT}{\mathcal{N}}
\newcommand{\Expr}{\mathcal{X}}
\newcommand{\Lang}{\mathcal{L}}

\newcommand{\str}{\mathrm{s}}
\newcommand{\chs}[1]{\mathtt{#1}}
\newcommand{\eos}{\lozenge}
\newcommand{\eosp}{\mathrm{\textit{\$}}}

\newcommand{\g}[1]{\mathsf{#1}}
\newcommand{\of}[2]{#1 \circ #2}
\newcommand{\upd}{up}
\newcommand{\ngs}{ngs}
\newcommand{\anyc}{\bullet}
\newcommand{\genseq}{\alpha\beta[\beta_1,\cdots,\beta_k,\beta_\emptyset,\g{B}_1,\cdots,\g{B}_k,\g{B}_\emptyset,l_1,\cdots,l_k,m]}

\newcommand{\ie}{\textit{i}.\textit{e}.}
\newcommand{\eg}{\textit{e}.\textit{g}.}

\newcommand{\etal}{\textit{et~al}. }

\newcommand{\lbl}[0]{\footnotesize}

\makeatletter
\newenvironment{mcases}[1][l]
 {\let\@ifnextchar\new@ifnextchar
  \left\lbrace
  \array{@{}l@{\quad}#1@{}}}
 {\endarray\right.}
\makeatother

\title{Derivatives of Parsing Expression Grammars}
\author{Aaron Moss
\institute{Cheriton School of Computer Science\\
University of Waterloo\\
Waterloo, Ontario, Canada}
\email{a3moss@uwaterloo.ca}}
\def\titlerunning{Derivatives of Parsing Expression Grammars}
\def\authorrunning{A. Moss}

\begin{document}

\maketitle

\begin{abstract}
This paper introduces a new derivative parsing algorithm for recognition of parsing expression grammars. 
Derivative parsing is shown to have a polynomial worst-case time bound, an improvement on the exponential bound of the recursive descent algorithm. 
This work also introduces asymptotic analysis based on inputs with a constant bound on both grammar nesting depth and number of backtracking choices; derivative and recursive descent parsing are shown to run in linear time and constant space on this useful class of inputs, with both the theoretical bounds and the reasonability of the input class validated empirically.
This common-case constant memory usage of derivative parsing is an improvement on the linear space required by the packrat algorithm.
\end{abstract}

\section{Introduction}

Parsing expression grammars (PEGs) are a parsing formalism introduced by Ford \cite{For02}. 
Any $\mathrm{LR}(k)$ language can be represented as a PEG \cite{For04}, but there are some non-context-free languages that may also be represented as PEGs (\eg{} $a^n b^n c^n$ \cite{For04}). 
Unlike context-free grammars (CFGs), PEGs are unambiguous, admitting no more than one parse tree for any grammar and input. 
PEGs are a formalization of recursive descent parsers allowing limited backtracking and infinite lookahead; a string in the language of a PEG can be recognized in exponential time and linear space using a recursive descent algorithm, or linear time and space using the memoized \emph{packrat} algorithm \cite{For02}. 
PEGs are formally defined and these algorithms outlined in Section~\ref{background-sec}. 
This paper introduces a PEG recognition algorithm in Section~\ref{algo-sec} based on the context-free grammar derivative of Might~\etal  \cite{MDS11}, which is in turn based on Brzozowski's regular expression derivative \cite{Brz64}. 
The essential idea of the derivative parsing algorithm is to keep an abstract syntax tree representing the ``rest'' of the original grammar, iteratively transforming it into a new grammar for every character in the input string. 
Unlike a packrat parser, which stores results of earlier parses to enable backtracking, the derivative parser never backtracks, instead calculating all possible ``backtracks'' concurrently at each point in the input. 
The analysis of this algorithm, presented in Section~\ref{bounds-sec}, reveals that it has quartic worst-case time complexity and cubic worst-case space complexity, though for inputs where both the grammar nesting depth and the number of backtracking options are bounded by a constant these bounds can be improved to linear time and constant space. 
This analysis is extended to consider the existing recursive descent and packrat algorithms on such inputs, providing a theoretical explanation for the superior practical performance of recursive descent, contrary to packrat's better asymptotic bounds.
The experimental results presented in Section~\ref{experimental-sec} empirically validate these theoretical predictions, as well as demonstrating the reasonability of the constant bounds for real-world inputs such as source code and structured data formats.

\section{Related work}

Ford \cite{For02} formalized PEG parsing and introduced the recursive descent and packrat algorithms. 
Medeiros and Ierusalimschy \cite{MI08} have developed a parsing machine for PEGs.
Their parsing machine is conceptually similar to a recursive descent parser, but they demonstrate superior performance in practice.

Mizushima~\etal \cite{MMY10} augment packrat parsers with a \emph{cut operator} which indicates that the current expression will not backtrack to another alternative. 
If there are no other valid alternatives for any expression currently being parsed then the entire packrat memoization table can be safely discarded; their empirical results on Java, XML, and JSON grammars show roughly constant space usage.
Mizushima~\etal{} also present a method for safe automatic insertion of cut operators, but their results demonstrate that it is not consistently effective and manual augmentation of the grammar is necessary for optimal performance.
Redziejowkski \cite{Red16} suggests that such automatic cut point insertion is likely to be insufficient for all but $\mathrm{LL}(1)$ grammars, limiting the utility of this approach.
The derivative parsing algorithm presented in this paper, by contrast, has no need for manual augmentation of the grammar, and can trim backtracking choices dynamically based on the input string, rather than only statically based on the grammar.

Kuramitsu \cite{Kur15} has developed an \emph{elastic sliding window} variant of packrat which uses constant space.
His method uses a fixed-size memoization table with hash collision-based eviction, relying on dynamic analysis to selectively stop memoizing some nonterminals.
Though reported runtime results using the elastic sliding window method are competitive with other approaches, it does forfeit the linear worst-case guarantee of packrat parsing.
Redziejowkski \cite{Red07} employs a similar but less sophisticated scheme in his Mouse parser-generator, storing a small fixed number of recent parsing results for each nonterminal. 
These methods heuristically discard backtracking information, in contrast to the derivative parsing algorithm, which keeps precisely the necessary backtracking information required.

Henglein and Rasmussen \cite{HR17} present \emph{progressive tabular parsing} (PTP) for the TDPL formalism PEGs are based on. 
PTP can iteratively generate a subset of the full parse table for a grammar on an input string, dynamically trimming a prefix of that table based on a tunable lookahead parameter. 
They prove linear time and space bounds for PTP; however, empirical comparison of their algorithm with other approaches is left to future work, so it is impossible to compare the constant factors to packrat, which has identical asymptotic bounds.
Henglein and Rasmussen also show some evidence that constant space usage may be achieved in practice for JSON by their dynamic table trimming mechanism; it would be interesting to compare the efficacy of their method to that of derivative parsing or the automatic table trimming approaches mentioned above on a wider variety of grammars.

Might~\etal \cite{MDS11} extended the Brzozowski regular expression derivative \cite{Brz64} to context-free grammars; Adams~\etal \cite{AHM16} further optimized this approach and proved a cubic worst-case time bound. 
The derivative parsing algorithm presented here is based on their work, with extensions to account for the lookahead and ordered choice features of PEGs.
Danielsson \cite{Dan10} uses derivatives as the underlying parsing mechanism of a parser combinator system.
Brachth\"{a}user~\etal \cite{BRO16} have developed a parser combinator encapsulating the derivative calculation, allowing parsers to be manipulated as first-class objects in a parser combinator framework; they use this method to compose distinct parsers in an interleaved manner.

\section{Parsing expression grammars}
\label{background-sec}

A PEG is expressed as a set of matching rules $A := \alpha$, where each \emph{nonterminal} $A$ is replaced by the \emph{parsing expression} $\alpha$. 
A parsing expression either matches an input string, possibly consuming input, or fails, consuming no input. 
Formally, given a set of strings $\Sigma^*$ over an alphabet $\Sigma$, a parsing expression $\alpha$ is a function $\alpha: \Sigma^* \rightarrow \left\lbrace \match, \fail \right\rbrace \times \Sigma^*$. 
$\NonT$ denotes the set of nonterminals, $\Expr$ the set of parsing expressions, and $\Rule$ the function mapping each nonterminal to its corresponding parsing expression, $\Rule: \NonT \rightarrow \Expr$. 
A grammar $\mathcal{G}$ is the tuple $(\NonT, \Expr, \Sigma, \Rule, \sigma)$, where $\sigma \in \Expr$ is the \emph{start expression} to parse. 
In this paper, lowercase Greek letters are used for parsing expressions ($\varphi \in \Expr$), uppercase Roman letters for nonterminals ($N \in \NonT$), lowercase monospace letters for characters ($\chs{c} \in \Sigma$), and lowercase non-italic Roman letters for strings ($\str \in \Sigma^*$).
For a parsing expression $\varphi$, the language $\Lang(\varphi)$ is the set of strings matched by $\varphi$, formally $\Lang(\varphi) = \left\lbrace \str \in \Sigma^* : \exists \str', \varphi(\str) = \left(\match, \str'\right) \right\rbrace$. 
A parsing expression $\varphi$ is \emph{nullable} if $\Lang(\varphi) = \Sigma^*$. 
Since parsing expressions match prefixes of their input, nullable expressions can be thought of as those that match the empty string without any restrictions on what comes after it.

The simplest parsing expression is the \emph{empty expression} $\varepsilon$; $\varepsilon$ always matches, consuming no input. 
Another fundamental parsing expression is the \emph{character literal} $c$ that matches and consumes a \texttt{c} or fails. 
Expressions are concatenated by a \emph{sequence expression}: $\alpha\beta$ matches if both $\alpha$ and $\beta$ match one after the other, failing otherwise. 
PEGs also provide an \emph{alternation expression}: $\alpha/\beta$ first attempts to match $\alpha$, then tries $\beta$ if $\alpha$ fails. 
Unlike the unordered choice in CFGs, once a subexpression matches in a PEG none of the others are tried. 
Specifically, $ab/a$ and $a/ab$ are distinct expressions, and the $ab$ in the second expression never matches, as any string it would match has already matched $a$. 
The \emph{negative lookahead expression} $!\alpha$ never consumes any input, matching if $\alpha$ fails. 
Expressions can be recursive; the \emph{nonterminal expression} $N$ matches only if its corresponding parsing expression $\Rule(N)$ matches, consuming whatever input $\Rule(N)$ does. 
For instance, $A := a\:A\;/\;\varepsilon$ matches a string of any number of \texttt{a}'s.

This repetition can be achieved by the \emph{repetition expression} $\alpha*$, syntactic sugar for a nonterminal like $A$ for $a*$ above. Due to the ordered choice in PEGs this matches greedily, so, \eg{} ${a*}~a$ never matches, since $a*$ consumes all available \texttt{a}'s, leaving none for $a$ to match. 
Since repetition expressions add no expressive power to PEGs this paper generally omits them from discussion.
Some other common syntactic sugar for PEGs is $\alpha?$ for $\alpha/\varepsilon$ and $\alpha+$ for $\alpha\,{\alpha*}$.
$\&\alpha$, the \emph{positive lookahead operator}, matches only if $\alpha$ matches, but consumes no input; it can be written $!(!\alpha)$. 
$\anyc$ abbreviates ``any character in $\Sigma$'', while $[a_0 \cdots a_k]$ is shorthand for the alternation $a_0/\cdots/a_k$. 
Less commonly, several uses of negative lookahead may be profitably simplified through use of a dedicated parsing expression: the negative character class $[^\wedge a_0 \cdots a_k]$ for $![a_0 \cdots a_k]\:\anyc$, the end-of-input expression $\eosp$ for $!\anyc$, and the \emph{until expression} $\alpha \rightarrow \beta$ for $(!\beta\,\alpha)\!*\,\beta$ (this repetition of $\alpha$ until a $\beta$ is seen is actually implemented as a fresh nonterminal $U := \beta/\alpha U$, taking advantage of the ordered choice property of PEGs to avoid matching $\beta$ twice).
Table~\ref{expr-table} has formal definitions of each of the fundamental expressions in decreasing order of precedence.

\begin{table}
\centering
\begin{align*}
a(\str)            & = \begin{cases} (\match, \str') & \str = \chs{a}\,\str' \\
                                     (\fail, \str)   & \text{otherwise} \end{cases} \\
\varepsilon(\str)  & = (\match, \str) \\
A(\str)            & = (\Rule(A))(\str) \\
!\alpha(\str)      & = \begin{cases} (\match, \str) & \alpha(\str) = (\fail, \str) \\
                                     (\fail, \str)  & \text{otherwise} \end{cases} \\
\alpha\beta(\str)  & = \begin{cases} (\match, \str'') & \alpha(\str) = (\match, \str') \wedge \beta(\str') = (\match, \str'') \\
                                     (\fail, \str)    & \text{otherwise} \end{cases} \\
\alpha/\beta(\str) & = \begin{cases} (\match, \str')  & \alpha(\str) = (\match, \str') \\
                                     (\match, \str'') & \alpha(\str) = (\fail, \str) \wedge \beta(\str) = (\match, \str'') \\
                                     (\fail, \str)    & \text{otherwise} \end{cases}
\end{align*}
\caption[Expression definitions]{Formal definitions of parsing expressions.}
\label{expr-table}
\end{table}

\subsection{Recursive descent algorithm}

Since parsing expression grammars are a formalization of recursive descent parsers, it should not be surprising that they may be recognized by a direct conversion of the expressions in Table~\ref{expr-table} into a recursive algorithm with a function for each nonterminal. 
This approach is simple and space-efficient, requiring only enough space to store the input string and function call stack (possibly linear in the size of the input). 
Recursive descent parsing may require exponential runtime; for example, consider a string of the form $\chs{a}^n\chs{c}^n$ as input to the grammar in Figure~\ref{exp-grammar}. 
In this case, the function $A$ is called twice after the first \texttt{a}, each of those calls resulting in two calls to $A$ after the second \texttt{a}, and so forth, as at each position $0 < i < n$ in the input string $A$ assumes that each of the $n-i$ remaining \texttt{a}'s is eventually followed by a \texttt{b}, fails, and must repeat this process assuming the \texttt{a}'s are followed by a \texttt{c}.

\begin{figure}[b]
\centering
\begin{align*}
S := & ~A~\eosp \\
A := & ~a~A~b~/~a~A~c~/~\varepsilon 
\end{align*}
\caption[Exponential-time grammar]{Grammar with exponential runtime using the recursive descent algorithm.}
\label{exp-grammar}
\end{figure}

This time bound does not account for left-recursive grammars, where a nonterminal may call itself without consuming any input. 
Though left-recursion is allowable in CFGs, it causes infinite loops in PEGs; an example of such a grammar would be $A := A\,a\:/\:\varepsilon$. 
Ford \cite{For04} defines a \emph{well-formed} grammar as one that contains no directly or indirectly left-recursive rules. 
This property is structurally checkable, so this paper considers left-recursion in PEGs to be undefined behaviour, and implicitly assumes all grammars are well-formed to guarantee a finite time bound for recursive descent and packrat parsing.

\subsection{Packrat parsing}
The essential idea of packrat parsing, introduced by Ford \cite{For02}, is to memoize calls to nonterminal functions in the recursive descent algorithm. 
If repetition expressions are treated as syntactic sugar for an anonymous nonterminal as described above, all parsing expressions except for nonterminal invocations can be parsed in time proportional to their number of subexpressions, a constant. 
Once a nonterminal has been evaluated at a position it can be parsed in constant time by reading the result from the memoization table. 
In any well-formed grammar, nonterminals do not left-recursively invoke themselves, and so may be parsed in constant time after parsing of all subsidiary nonterminals.
Since there are a constant number of nonterminals, which may be parsed no more than once for each position in the input, this algorithm takes linear time, using linear space to store the memoization entries for each nonterminal at each index.

\section{Derivative parsing}
\label{algo-sec}
This paper presents a new algorithm for recognizing parsing expression grammars based on the derivative parsing algorithm for context-free grammars introduced by Might~\etal \cite{MDS11}. 
Essentially, the \emph{derivative} of a parsing expression $\varphi$ with respect to a character \texttt{c} is a new parsing expression $d_c(\varphi)$ that recognizes everything after the initial \texttt{c} in $\Lang(\varphi)$. 
As an example, $d_f\left(foo/bar/baz\right) = oo$, while $d_b\left(foo/bar/baz\right) = ar/az$, $d_a\left(ar/az\right) = r/z$, and $d_r\left(r/z\right) = \varepsilon$.
Where the specific character is not relevant, the derivative of $\varphi$ may be written $\varphi'$. 
Formally, $\Lang(d_c(\varphi)) = \left\lbrace \str : \chs{c}\,\str \in \Lang(\varphi) \right\rbrace$. 
The essence of the derivative parsing algorithm is to take repeated derivatives of the original parsing expression for each character in the input and to check for a nullable or failure parser at each step. 
The repeated derivative $d_{c_k} \circ \cdots \circ d_{c_1}(\varphi)$ may be written $d_{c_1 \cdots c_k}(\varphi)$, or $\varphi^{(k)}$ where the specific characters are not important.
To compute derivatives of parsing expressions, two new parsing expressions are introduced for failure states: $\varnothing$ and $\infty$; $\varnothing(\str) = \infty(\str) = \left(\fail, \str\right)$. 
$\varnothing$ represents a generic parse failure, while $\infty$ represents a left-recursive infinite loop, which this algorithm can detect and terminate on. 

Derivative parsing essentially tries all possible parses concurrently; since parsing expression grammars are inherently unambiguous, they may actually be better suited to this approach than context-free grammars. 
Nonetheless, tracking exactly which expressions in the current parse tree can match against or consume the current character is the key problem in this algorithm. 
To solve this problem, the algorithm uses the concept of \emph{backtracking generations} (\emph{generations}) as a way to account for choices in the parsing process. 
Each backtracking generation represents a point in the input at which a parsing expression may have matched and ceased to consume input.
Generation $0$ is the ``leftmost'' ongoing parse, where each new character is consumed by the first subexpression of any sequence expression, while higher generations represent backtracking decisions that occurred at some earlier point and have not yet been resolved.
Expressions may have multiple outstanding backtracking decisions; to represent this, a set of generations is denoted with uppercase sans-serif characters, such that $\g{X} = \{x_0, x_1, \cdots, x_k\}$, where the indices $0, \cdots, k$ are assigned from smallest to largest value.

To perform the derivative computation, some expressions must be augmented with generation information. 
The $\varepsilon$ expression is augmented with a generation $i$, written $\varepsilon[i]$; $\varepsilon[0]$ represents a match at the current input location, while $\varepsilon[i > 0]$ means that the backtracking decision deferred at generation $i$ did actually result in a match. 
In general, expressions may have different generation sets than their subexpressions; to translate one generation set into another composition of generation sets is defined as $\of{\g{F}}{\g{G}} = \{f_i : i \in \g{G}\}$ (\eg\ $\of{\{0, 3, 4\}}{\{0, 2\}} = \{0, 4\}$). 
Alternation expressions $\alpha/\beta$ are annotated with $[\g{A},\g{B},m]$, where $\g{A}$ and $\g{B}$ are the mapping sets for $\alpha$ and $\beta$, respectively, and $m$ is the maximum backtracking generation for the expression.
$\g{A}$ and $\g{B}$ map new generations of $\alpha$ and $\beta$ from the same input index to the same value, and generations that first occur at different input indices to different values, allowing the separate backtracking generations of $\alpha$ and $\beta$ to be expressed as generations of the compound expression $\alpha/\beta$.
As an example, consider $\varphi = a!c/!(ab)$; $a!c$ starts by consuming characters, so its generation mapping set $\g{A} = \{0\}$, while $!(ab)$ represents a \emph{possible} match (which would later produce a $\varepsilon[1]$), so its initial mapping set $\g{B} = \{0, 1\}$, to map that $\varepsilon[1]$ to a generation $1$ match of the containing expression. 
Thus, the initial form of $\varphi$ is $a!c/!(ab)[\{0\},\{0,1\},1]$. 
When the derivative of this expression is taken, the generation information must be updated; $d_a(\varphi) = !c/!b$, but since the $!c$ represents a possible match at a later position than the $!b$ it needs a new backtracking generation, $2$.
To map a $\varepsilon[1]$ success result from $!c$ to a $\varepsilon[2]$ for $\varphi$, the generation information is updated to $!c/!b[\{0,2\},\{0,1\},2]$.

Since sequence expressions $\alpha\beta$ encode the successor relationship, the bulk of the complication of dealing with backtracking must also be handled in sequence expressions. 
The first type of backtracking handled by a sequence expression is \emph{nullability backtracking}, where $\alpha$ is a nullable but non-empty expression, \eg\ $\alpha\beta = (\delta/\varepsilon)\beta$. 
This example is almost equivalent to $\delta\beta/\beta$, and is handled similarly to an alternation expression, except that a later nullable derivative $\delta^{(k)}$ causes the second alternative $\beta^{(k)}$ to be discarded\footnotemark. 
To account for this possibility the sequence expression $\alpha\beta$ is augmented with $\beta_\emptyset$ and $\g{B}_\emptyset$, the \emph{failure backtrack} and the generation set that maps its backtracking generations into the sequence expression's generation space. 
$\beta_\emptyset$ is the parse to continue if $\alpha^{(k)}$ is nullable (\ie\ a match, where $\beta$ should possibly begin parsing) but some later $\alpha^{(k+i)}$ fails (\ie\ $\alpha^{(k)}$ was the match which should be continued from).

\footnotetext{$(\delta/\varepsilon)\beta$ is not exactly equivalent to $\delta\beta/\beta$; consider $\delta = a$, $\beta = a+$; $\chs{a} \in \Lang(a\:{a+}/{a+})$, but $\chs{a} \not \in \Lang((a/\varepsilon){a+})$, as $(a/\varepsilon)$ consumes the $\chs{a}$ but $a+$ still requires one.}

Sequence expressions $\alpha\beta$ also handle \emph{lookahead backtracking}. 
For each generation where $\alpha$ may have ceased to consume input, yet did not unambiguously match or fail (\eg\ by including a lookahead expression), $\alpha\beta$ must begin to track derivatives of $\beta$ so that the proper derivative can be computed if one of those generations eventually results in a match. 
Sequence expressions implement lookahead backtracking by keeping the \emph{lookahead follower} $\beta_i$, the derivative of $\beta$ started when generation $i$ first appeared, its associated generation map $\g{B}_i$, and the index $l_i$ of the last generation of $\alpha\beta$ where $\beta_i$ was nullable ($0$ for none such). 
The $l_i$ are necessary to track possible matches of $\alpha\beta$ where both subexpressions have been consumed. 
In addition to nullability and lookahead backtracking information, sequence expressions are also annotated with $m$, the maximum backtracking generation of the expression.

\begin{table}
\centering
\begin{tabular}{r|l|l}
$\varphi$                     & $back(\varphi)$            & $match(\varphi)$   \\
\hline
$a$                           & $\{0\}$                    & $\{\}$             \\
$\varepsilon[i]$              & $\{i\}$                    & $\{i\}$            \\
$\varnothing$                 & $\{0\}$                    & $\{\}$             \\
$\infty$                      & $\{0\}$                    & $\{\}$             \\
$A$                           & $back(\Rule(A))$ where $back(A) = \{0\}$ 
                              & $match(\Rule(A))$ where $match(A) = \{\}$ \\
$!\alpha$                     & $\{1\}$                    & $\{\}$             \\
$\alpha[\g{A},m]$             & $\of{\g{A}}{back(\alpha)}$ & $\of{\g{A}}{match(\alpha)}$    \\
$\alpha/\beta[\g{A},\g{B},m]$ & $\of{\g{A}}{back(\alpha)} \cup \of{\g{B}}{back(\beta)}$ 
                              & $\of{\g{A}}{match(\alpha)} \cup \of{\g{B}}{match(\beta)}$ \\
\end{tabular}

\begin{align*}
\varphi        = &~\genseq \\
back(\varphi)  = &~(0 \in back(\alpha)~?~\{0\} : \{\}) 
                   \cup_{i \in 1..k} (\of{\g{B}_i}{back(\beta_i)}) \cup \{l_i : l_i > 0\}
                   \cup \of{\g{B}_\emptyset}{back(\beta_\emptyset)} \\
match(\varphi) = &~\cup_{i \in match(\alpha)} (\of{\g{B}_i}{match(\beta_i)}) \cup \{l_i : l_i > 0\} 
                   \cup~\of{\g{B}_\emptyset}{match(\beta_\emptyset)}
\end{align*}
\caption{$back$ and $match$ Definitions}
\label{back-match-table}
\end{table}

More formally, the functions $back$ and $match$ produce sets of backtracking generations from parsing expressions; $back(\varphi)$ is the set of backtracking generations currently being parsed by $\varphi$ while $match(\varphi)$ is the subset of $back(\varphi)$ that has successfully matched. In terms of the existing literature on derivative parsing, $match$ is an extension of the nullability function and $0 \in match(\varphi)$ means that $\varphi$ is nullable. 
The values of $back$ and $match$ for all expressions are defined in Table~\ref{back-match-table}; parsing expressions which need additional bookkeeping information to perform these calculations include it in square brackets. 
Note that $back$ and $match$ for a nonterminal $A$ are fixed-point computations where $A$ acts like an infinite loop $\infty$ if encountered recursively. 
This yields correct results if $match$ is pre-computed for each subexpression in the original grammar by a fixed-point computation. 
Might~\etal \cite{MDS11} use an approach based on Kleene's least-fixed-point theorem that can be straightforwardly adapted to this algorithm. 
Once this initial fixed point is calculated the nonterminal rules preclude the introduction of any further recursion, and thus the need for further fixed point calculations. 
A \emph{map expression} $\alpha[\g{A},m]$ is also introduced here to remap $back$ and $match$ sets into a different generation space; it has the same semantics as $\alpha$, \ie\ $\alpha[\g{A},m](\str) = \alpha(\str)$.

Derivatives update the generation maps of their expressions via the following update function: 
\[\upd(\g{P},\varphi,\varphi',m) = \begin{cases} \g{P} \cup \{m+1\} & \max(back(\varphi')) > \max(back(\varphi)) \\
                                                 \g{P}              & \text{otherwise} \end{cases}\]

$up$ adds a new generation to the map $\g{P}$ for any expression $\varphi$ that adds a new backtracking generation when transformed to $\varphi'$. 
Since backtracking generations correspond to points in the input where a backtracking decision has been deferred, there is no more than one new generation for each parsing expression per derivative taken, and expressions statically defined in the original grammar can have a backtracking generation no greater than $1$. 
To transform these static generations into the generation space of an ongoing derivative computation, the following \emph{new generation set} function is defined:  
\[\ngs(\varphi,m) = \begin{cases} \{0,m+1\} & \max(back(\varphi)) > 0 \\
                                  \{0\}     & \text{otherwise} \end{cases}\]

Before it can be used in a derivative calculation, the original grammar needs some pre-processing to insert backtracking annotations. 
Any $\varepsilon$ expressions in the original grammar become $\varepsilon[0]$. 
For an alternation expression $\alpha/\beta[\g{A}, \g{B}, m]$ the initial values of $\g{A}$ and $\g{B}$ are $back(\alpha)$ and $back(\beta)$, respectively; $m$ is initialized to $\max(\g{A}, \g{B})$. 
Sequence expressions are more complicated: given a sequence expression $\alpha\beta$, if $\alpha$ is nullable (\ie{} $0 \in match(\alpha)$) then the initial failure lookahead $\beta_\emptyset = \beta$, otherwise $\beta_\emptyset = \varnothing$; in either case $\g{B}_\emptyset = back(\beta_\emptyset)$. 
If $1 \in back(\alpha)$ then $\alpha\beta$ has an initial lookahead follower $\beta_1 = \beta$ with a generation set $\g{B}_1 = back(\beta)$, if $\beta$ is nullable then the lookahead nullability marker $l_1 = 1$, otherwise $l_1 = 0$; $m = \max(0, \g{B}_\emptyset, \g{B}_1, l_1)$, for any of these defined.
All these constructions preserve the property that expressions in the original grammar have a maximum backtracking generation of $1$.

\begin{table*}
\centering
\begin{subfigure}[t]{\textwidth}
\begin{align*}
a'                           & = \begin{cases} \varepsilon[0] & c = \chs{a} \\
                                               \varnothing    & \text{otherwise} \end{cases} &
\varepsilon[i]'              & = \begin{cases} \varepsilon[i] & i > 0 \lor c = \eos \\
                                               \varnothing    & i = 0 \end{cases} \\
\varnothing'                 & = \varnothing &
\infty'                      & = \infty
\end{align*}
\end{subfigure}
\begin{subfigure}[t]{\textwidth}
\begin{align*}
A'                           & = \Rule(A)' \text{ under } A' = \infty \\
(!\alpha)'                   & = \begin{cases} \varnothing    & match(\alpha') \neq \emptyset \\
                                               \varepsilon[1] & \alpha' = \varnothing \\
                                               \infty         & \alpha' = \infty \\
                                               !(\alpha')     & \text{otherwise} \end{cases} \\
\alpha[\g{A},m]'             & = \begin{mcases}[lrl] \varepsilon[\of{\g{A}'}{\{i\}}] & \alpha' = \varepsilon[i] & \text{where}~\g{A}' & = \upd(\g{A},\alpha,\alpha',m) \\
                                                     \varnothing                     & \alpha' = \varnothing    &                  m' & = \max(m, \g{A}') \\
                                                     \infty                          & \alpha' = \infty & & \\
                                                     \alpha'                         & \g{A}' = \{0,1,...,m'\} & & \\
                                                     \alpha'[\g{A}',m']              & \text{otherwise} & & \end{mcases} \\
\alpha/\beta[\g{A},\g{B},m]' & = \begin{mcases}[lrl] \beta'[\g{B}',m']  & \alpha' = \varnothing                                     & \text{where}~\g{A}' & = \upd(\g{A},\alpha,\alpha',m) \\
                                                     \infty             & \alpha' = \infty                                          &              \g{B}' & = \upd(\g{B},\beta,\beta',m) \\
                                                     \alpha'[\g{A}',m'] & \beta' = \varnothing \lor match(\alpha') \neq \emptyset   &                  m' & = \max(m,\g{A}',\g{B}') \\
                                                     \alpha'/\beta'[\g{A}',\g{B}',m'] & \text{otherwise} & & \end{mcases} \\
\mathrlap{\genseq'}~~~~~~~~~~~~~~~~~~~~~ \\
& = \begin{cases} \beta[\ngs(\beta,m),m']                        & \alpha' = \varepsilon[0] \\
                  \beta_i'[\g{B}_i',m']                             & \alpha' = \varepsilon[0 < i \leq k'] \land l_i' = 0 \\
                  (\beta_i'/\varepsilon[1])[\g{B}_i',\{0,l_i'\},m'] & \alpha' = \varepsilon[0 < i \leq k'] \land l_i' > 0 \\
                  \beta_\emptyset'[\g{B}_\emptyset',m']             & \alpha' = \varnothing \\
                  \infty                                            & \alpha' = \infty \\
                  \alpha'\beta^\dagger[\beta_1',\cdots,\beta_{k'}',\beta_\emptyset^\dagger,\g{B}_1',\cdots,\g{B}_{k'}',\g{B}_\emptyset^\dagger,l_1',\cdots,l_{k'}',m'] & \text{otherwise} \end{cases} \\
\text{where~~~~~~~~~~~} & \\
m'                              & = \max(m,\g{B}_i',\g{B}_\emptyset',l_i') \text{ for any of these used}
\end{align*}
\end{subfigure}
\begin{subfigure}[t]{\textwidth}
\begin{align*}
\beta^\dagger                   & = \begin{cases} \beta' & c = \eos \land \alpha' = \varepsilon[i \in \{0, k+1\}] \\
                                                  \beta  & \text{otherwise} \end{cases} &
k'                              & = \begin{cases} k+1 & \max(back(\alpha')) > \max(back(\alpha)) \\
                                                  k   & \text{otherwise} \end{cases} \\
\beta_{k+1}'                    & = \beta &
\beta_\emptyset^\dagger         & = \begin{cases} \beta            & 0 \in match(\alpha') \\
                                                  \beta_\emptyset' & \text{otherwise} \end{cases} \\
\begin{aligned}\g{B}_{1 \leq i \leq k}' \\ \g{B}_{k+1}'\end{aligned} &
\begin{aligned} = \upd(\g{B}_i,\beta_i,\beta_i',m) \\ = \ngs(\beta,m)~~~~~~~~~~\end{aligned} &
\g{B}_\emptyset^\dagger         & = \begin{cases} \ngs(\beta,m)                           & 0 \in match(\alpha') \\
                                                  \upd(\g{B}_\emptyset,\beta_\emptyset,\beta_\emptyset',m) & \text{otherwise} \end{cases} \\
l_{k+1}                         & = 0 &
l_{1 \leq i \leq k+1}'          & = \begin{cases} m+1 & 0 \in match(\beta_i') \\
                                                  l_i & \text{otherwise} \end{cases}
\end{align*}
\end{subfigure}
\caption[Expression derivatives]{Definition of $d_c(\varphi)$; $\delta'$ abbreviates $d_c(\delta)$.}
\label{deriv-table}
\end{table*}

Having defined the necessary helper functions, Table~\ref{deriv-table} describes a single derivative step.
$a'$ is a straightforward derivative of a terminal, consuming an $\chs{a}$ or failing. 
$\varepsilon[0]'$ is a failure on anything except end-of-input because there is nothing left in an $\varepsilon[0]$ to take a derivative of, while $\varepsilon[i > 0]'$ is preserved to indicate a possible match that has yet to be verified; the failure expressions $\varnothing$ and $\infty$ are similarly preserved by the derivative operation. 
Nonterminals are again assumed to be $\infty$ while calculating their derivatives to break infinite loops in the computation; for well-formed grammars this case should not occur. 
Negative lookahead, mapping and alternation expressions are basically distributive with respect to the derivative; the initial three cases for both $(!\alpha)'$ and $\alpha[\g{A},m]'$ distribute matches and failures according to the appropriate semantics, while the fourth case of $\alpha[\g{A},m]'$ strips unnecessary identity maps. 
The third case for $\alpha/\beta[\g{A},\g{B},m]'$ is of particular note; it implements the ordered choice property of PEGs by stripping the $\beta$ alternative in the case of any match in the $\alpha$ alternative.

The first case of the sequence expression derivative accounts for a straightforward match of $\alpha$ at generation 0, which is followed by a parse of $\beta$ mapped into the sequence expression's generation space. 
In the second and third cases $\alpha$ has matched at generation $i > 0$ and is mapped to the appropriate lookahead follower; the third case accounts for the possibility that the lookahead follower has already passed its final match by alternating it with a match at its nullability index $l_i$. 
The fifth case propagates an infinite loop failure, while the fourth case produces the failure backtrack if $\alpha$ fails; this backtrack will be a failure expression itself if $\alpha$ has never matched. 
Generally speaking, the sixth case takes derivatives of $\alpha$ and all the $\beta_i$ and $\beta_\emptyset$ lookahead and nullability backtracks. 
Particularly, $k'$ accounts for whether $\alpha$ produced any new backtracking generations in this derivative step, while $l_i'$ tracks the last nullability of the lookahead backtracks. 
The $\beta_\emptyset^\dagger$ and $\g{B}_\emptyset'$ cases are especially notable, as they implement the ordered choice property of PEGs by resetting to $\beta$ whenever $\alpha$ is nullable. 
The special cases for end-of-input ($c = \eos$) are necessary to resolve outstanding lookahead expressions; for instance, $!\anyc$ is one idiomatic way to represent end-of-input in a PEG, but $\anyc$ must parse to a failure by receiving a non-character (\ie\ $\eos$) to recognize $!\anyc$ as a match. 
In the $\beta^\dagger$ special case of the sequence expression derivative, $\beta$ must also be matched against $\eos$ if $d_\eos(\alpha)$ resulted in a fresh $\varepsilon$-expression.

Given this definition of derivative, the $parse$ function defined in Table~\ref{deriv-fn-table} evaluates a parsing expression $\varphi$ on an input string $\chs{c_1 c_2 \ldots c_n}$. 
This function computes repeated derivatives until either the input is exhausted or the derivative becomes a success or failure value. 
The fact that the rest of the string is not returned on a match should not be a practical problem: further parsing could be performed by wrapping the start expression in another expression or the $parse$ function could also be augmented with logic to track the indices of each backtracking generation to return the proper suffix of the input on a match.

\begin{table}
\centering
\begin{align*}
parse(\varphi,\str = [\chs{c_1 c_2 \ldots c_n} \eos]) = \begin{cases} 
    \match                                                  & match(\varphi) \neq \emptyset \\
    \fail                                                   & \varphi \in \{\varnothing, \infty\} \\
    \fail                                                   & \str = [~] \land match(\varphi) = \emptyset \\
    parse(d_{c_1}(\varphi), [\chs{c_2 \ldots c_n} \eos]) & \text{otherwise}
\end{cases}
\end{align*}
\caption{Derivative parsing recognizer function}
\label{deriv-fn-table}
\end{table}

The derivative parsing algorithm as presented works correctly, but has exponential time and space complexity, since each subexpression of a parsing expression may expand by the size of the grammar during each derivative. 
This problem can be avoided by memoizing the derivative computation; if a table of parsing expression derivatives $\varphi'$ is constructed for each derivative step, duplicate derivative calculations can be avoided by using a single instance of $\varphi'$ for all derivatives of $\varphi$.
This changes the derivative tree into a directed acyclic graph (DAG), and also bounds the increase in derivative size and computation time, allowing this parsing algorithm to achieve the polynomial time and space bounds proved in Section~\ref{bounds-sec}. 
This memoization is also the reason for the generation sets and $back$ and $match$ functions; these mechanisms allow a single subexpression to have multiple parents, yet be treated differently in each context. 
As an additional optimization, the values of $back$ and $match$ are also memoized.

The treatment of backtracking generations can also be improved by eliminating information that is never used. 
If a generation $i$ is not present in $back(\alpha)$, then the sequence expression $\alpha\beta$ does not need to keep $\beta_i$, $\g{B}_i$, and $l_i$, since $\alpha$ can never reduce to $\varepsilon[i]$, which would select $\beta_i$. 
Similarly, in the map expression $\alpha[\g{A} = \{a_0,a_1,\cdots,a_k\},m]$, it is not necessary to store $a_i$ for any $i \not\in back(\alpha)$, so long as the indices for the other values in $\g{A}$ are preserved; this transforms $\g{A}$ to a partial function of indices to values. 
The same principle can be applied to the backtracking sets in the alternation and sequence expressions as well.
If these optimizations are applied, the space required for backtracking information becomes proportional to the number of current backtracking choices, rather than the potentially much larger number of backtracking choices seen throughout the parsing process.

To summarize, the derivative parsing algorithm for PEGs works by iterative transformation of an abstract syntax tree representing the ``rest'' of the grammar. 
Unlike a packrat parser, which stores parse results for efficient backtracking, the derivative parser never backtracks, instead calculating all possible ``backtracks'' concurrently at each point in the input. 
Unlike CFGs, PEGs do not allow multiple parse trees or left-recursion, so this algorithm produces only one parse tree, and does not require general-purpose fixed-point computations after pre-processing, as in derivative parsing for CFGs.

\section{Analysis}
\label{bounds-sec}
The derivative parsing algorithm runs in time proportional to the cost of a derivative computation times the size of the input, and requires as much space as the largest derivative computed. 
The cost of a derivative computation $d_c(\varphi)$ is linear in the size of $\varphi$, but since $\varphi$ is a memoized DAG, it may re-use the computation of some sub-expressions of $\varphi$. 
To account for this, $^\#(\varphi)$ is defined to be the number of \emph{unique} subexpressions of $\varphi$, including $\varphi$ itself. 
Some expressions are larger than others, and take longer to compute the derivative of, \eg\ $\alpha/\beta[\g{A},\g{B},m]$ requires more storage and computation than $!\alpha$.
To formalize this idea, the asymptotic bounds on the size $|\varphi|$ of the root node of $\varphi$ are given in Table~\ref{size-table}; this size is expressed in terms of $b$, the current number of generations. 
Each backtrack set is obviously $O(b)$, while a sequence expression keeps bookkeeping information for $O(b)$ generations.

\begin{table}[b]
\centering
\begin{align*}
|a| = |\varepsilon[i]| = |\varnothing| = |\infty| = |A| = |!\alpha| & = O(1) \\
                  |\alpha[\g{A},m]| = |\alpha/\beta[\g{A},\g{B},m]| & = O(b) \\
                                                          |\genseq| & = O(b^2)
\end{align*}
\caption[Expression size]{Size $|\varphi|$ of the root node of an expression $\varphi$ in terms of $b$, the current number of generations.}
\label{size-table}
\end{table}

It is evident from Table~\ref{deriv-table} that, with the exception of $A'$, no derivative adds more expression nodes or makes an existing node asymptotically larger.
In fact, many rules strictly reduce the size of their expression node or discard subexpressions. Furthermore, since none of the derivative rules add more than one new backtracking generation, $b$ is linear in the number of derivatives taken. 

Having bounded the size of a single expression node, it can now be shown that the number of expression nodes is also linear in the number of derivatives taken.
The proof of this statement follows directly from the fact that taking the derivative of a nonterminal $A$ is the only operation that adds nodes to the expression, replacing $A$ with $\Rule(A)'$. 
But since $\Rule(A)$ is statically defined $^\#(\Rule(A)')$ is a constant, and since derivatives are memoized the expression nodes comprising $\Rule(A)'$ are only added once per derivative step. 
$\Rule(A)'$ may include expansions of other nonterminals to their respective derivatives, but cannot include a recursive expansion of $A$, since recursive expansions of $A$ are replaced by $\infty$. 
Since an $\infty$ expression is the same size as an $A$ expression, this expansion does not add any expression nodes, and $^\#(\Rule(A)')$ is bounded by a constant the size of the grammar.
According to this argument, the number of expression nodes added during a derivative computation is bounded by $\sum_{A \in \NonT} {^\#}(\Rule(A)')$, a constant, and thus the number of expression nodes is at most linear in the number of derivatives taken.

Having established that the number of expression nodes is at most linear in the number of derivatives taken and the size of each of those nodes is no more than quadratic in the number of backtracking options (where the number of backtracking options is also linear in the number of derivatives taken), it follows directly that the space required by the derivative parsing algorithm is $O(b^2 n)$ ($O(n^3)$). 
As each derivative operation must traverse one of these expressions, the worst-case running time for this algorithm is $O(b^2 n^2)$ ($O(n^4)$). 
This polynomial bound guarantees quicker worst-case execution than the recursive descent algorithm, but does not match the bound of packrat parsing.

\subsection{``Well-behaved'' inputs}
\label{well-behaved-sec}
However, many real-world inputs have properties that allow these worst-case bounds to be significantly tightened. 
If the number of backtracking choices $b$ at each point in the input is bounded by a constant then each expression node is also constant size and these bounds are reduced to linear space and quadratic time. 
The results in Section~\ref{experimental-sec} show this bound holds in practice; as a theoretical argument, many practical grammars are LR, and thus can choose the correct branch of an alternation expression after reading a constant number of terminals past the end of the expression.

It is well-known (and verified in Section~\ref{experimental-sec}) that real-world inputs such as source code and structured data typically do not contain deeply nested recursive structure.
To formalize this, the \emph{nesting depth} of an input string with respect to a grammar is the maximum number of nested expressions containing any terminal expression, where the depth of $\alpha$ in $\alpha*$ is $1$, as in an iterative or tail-call optimized implementation. 
Bounding the nesting depth of the input by a constant implies an equivalent constant bound on the depth of the DAG representing the derivative parser, as the DAG can only get deeper by taking the derivative of a nonterminal.
Since the bound on backtracking restricts fanout, there is a constant bound on the number of expression nodes produced in the derivative parse of an input file with constant-bounded nesting depth and backtracking choices.
Since these nodes require constant space, as shown above, these two conditions on the input are sufficient to guarantee constant space usage for a derivative parser, implying a linear time bound.

The time and space bounds of the recursive descent algorithm can also be tightened for inputs with constant-bounded backtracking and nesting depth.
Nesting depth corresponds directly to stack size for this algorithm, so a constant bound on nesting depth is also a constant bound on the space usage of the recursive descent algorithm, assuming the input is not counted in this space bound and allows random seek (otherwise recursive descent still needs linear space to buffer the input). 
For the recursive descent algorithm, a bound on the number of backtracking options $b$ at each point in the input can be interpreted as a bound on the number of times any character in the input is read; hence, a constant bound on $b$ implies that no character is read more than $b$ times, a linear bound on the runtime of the algorithm. 

It has been empirically shown \cite{Kur15,MMY10} that packrat parsers can also achieve linear time and constant space for many practical grammars by selectively discarding backtracking information; however, the two conditions discussed here are not sufficient to guarantee this space usage. 
As evidence, consider ${a*}\,b\:/\:{a*}\,c$; the memoization table should store entries for every \texttt{a} in the input (which each begin an $a*$ parse), a linear space requirement, while both the derivative and recursive descent algorithms would use constant space to parse this expression.
It is known \cite{BS08,Kur15,MI08} that recursive descent parsing often out-performs packrat parsing in practice, contrary to the worst-case analysis; to the author's knowledge, the explanation of this phenomenon as a class of practical inputs on which recursive descent has superior asymptotic bounds to packrat is a novel contribution of this work.

In summary, derivative parsing has a polynomial runtime bound, unlike recursive descent parsing, yet can parse a wide class of useful inputs in linear time and constant space, improving on the space bound of packrat parsing with no asymptotic sacrifice in time.

\section{Experimental results}
\label{experimental-sec}

The author has implemented a derivative parser in C++ to empirically investigate the runtime performance of derivative parsing. 
The most significant optimization not covered in the theoretical description of the algorithm is that the implementation takes advantage of the associativity of alternation to flatten binary trees of alternation nodes to lists of choices; this resulted in modest speedup for grammars containing alternations with many options. 
As done by Adams~\etal \cite{AHM16}, the implementation also uses a memoization field in each node instead of a separate table.
Limiting the number of lookahead expressions by replacing them with negative character classes, end-of-input, and until expressions was also a useful optimization, saving 20\% runtime in some tests.
An imperative variant of the algorithm was attempted, with the intent to increase reuse of long-lived compound expression nodes by utilizing mutation, but it did not increase performance and added significant implementation complexity. 

\begin{figure}[b]
\centering
\include{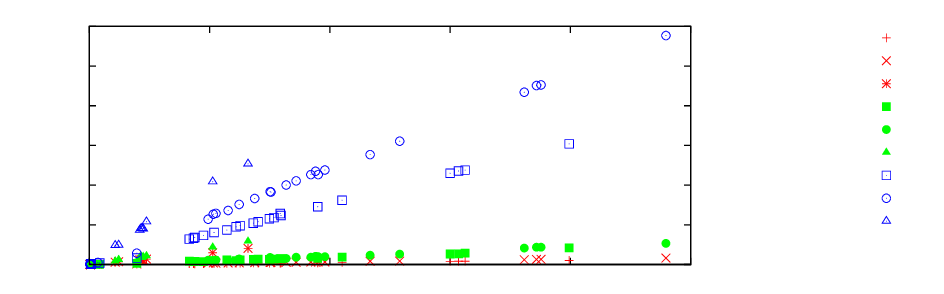}
\caption[Algorithm Runtime]{Algorithm runtime with respect to input size; lower is better.}
\label{runtime-fig}
\end{figure}

\begin{figure}
\centering
\include{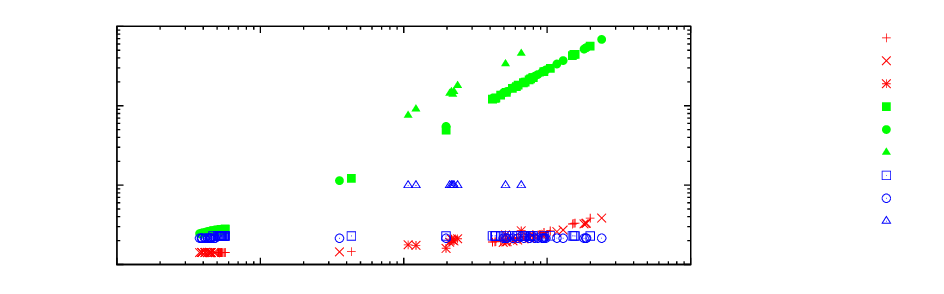}
\caption[Algorithm Memory Use]{Maximum algorithm memory use with respect to input size; lower is better.}
\label{mem-use-fig}
\end{figure}

\begin{figure}
\centering
\include{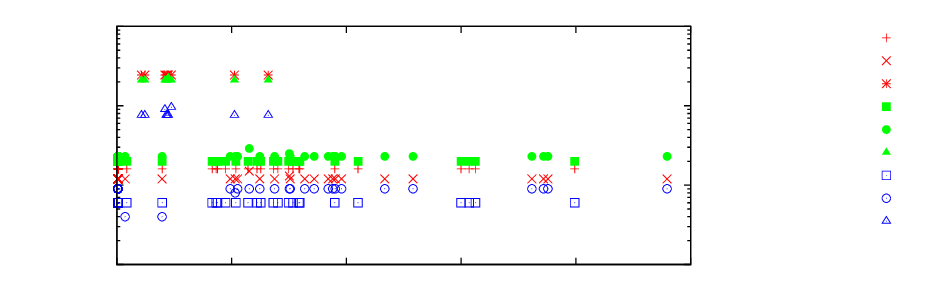}
\caption[Maximum Backtracking]{Maximum backtracking performed by each algorithm, plotted against input size.}
\label{backtracks-fig}
\end{figure}

\begin{figure}
\centering
\include{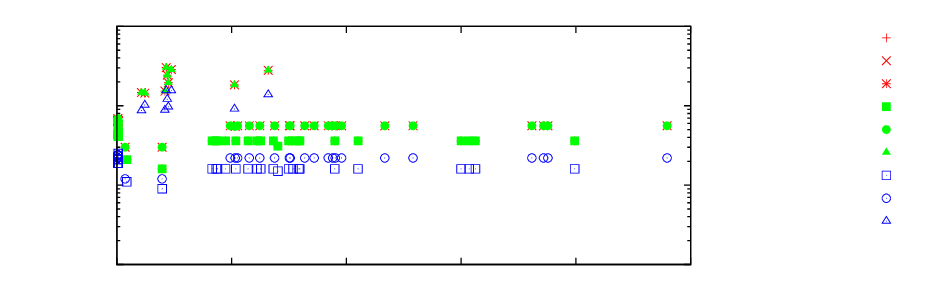}
\caption[Algorithm Memory Use]{Maximum nesting depth for each algorithm, plotted against input size.}
\label{nesting-depth-fig}
\end{figure}

To test the predictions made by the analysis presented in Section~\ref{bounds-sec}, parser-combinator-based recursive descent and packrat parser-generators were also implemented by the author. 
All three parsers were tested using XML, JSON, and Java inputs from Mizushima~\etal \cite{MMY10}. 
Code and test data are available online \cite{Mos14}.
All tests were compiled with g++ 6.2.0 running on a machine with 8 GB of RAM, a dual-core 2.6 GHz processor, and SSD main storage. 
No formal proof of the correctness of the derivative algorithm is presented, but it produces the same results as the other algorithms on an extensive test suite.

Figure~\ref{runtime-fig} shows the runtimes of each of the three algorithms, separated by grammar and graphed against input size. 
As can be seen, all three algorithms take linear time on the given inputs, though the constant factor depends both on the algorithm and the complexity of the grammar. 
More complex grammars result in longer runtime; in this case the XML grammar has 23 nonterminals, the JSON grammar has 24 nonterminals, and the Java grammar has 178 nonterminals. 

Figure~\ref{mem-use-fig} demonstrates that the linear time guarantee provided by packrat parsing comes at a heavy cost in memory (note the logarithmic scale on both axes), with the linear bound on memory usage fitting tightly and resulting in memory usage two orders of magnitude greater than the size of the input. 
By contrast, derivative parsing uses constant space comparable to the input size for the well-behaved instances in this test, while the recursive descent algorithm uses a similarly small amount of space, with the slight linear growth due to buffering of the input to facilitate backtracking.

Figures~\ref{backtracks-fig} and~\ref{nesting-depth-fig} empirically validate the claims in Section~\ref{well-behaved-sec} of constant bounds on backtracking and nesting depth. 
Backtracks reported in Figure~\ref{backtracks-fig} are the maximum number of times the input string or memoization table was read at any one index for recursive descent and packrat parsing, and the maximum number of leaf nodes in the expression tree for derivative parsing; it can be seen from these results that these maximums are effectively constant for a given grammar.
It is interesting to note that packrat memoization is not markedly effective at reducing backtracking, with backtracks eliminated by memoization fairly closely balanced by duplicated hits due to memoization cache misses.
The lower amount of backtracking in the derivative algorithm is likely an artifact of the methodology: since failures are removed as part of the derivative function their nodes are not counted by the instrumentation.
Figure~\ref{nesting-depth-fig} shows roughly constant maximum nesting depth for each grammar; the recursive and packrat algorithms have essentially identical results, as should be expected, while the derivative algorithm's lesser nesting depth is due to dynamic compaction of the derivative DAG based on the input -- if one branch of an alternation fails the alternation will be replaced with its successful branch in the derivative algorithm, while the alternation expression will still be on the recursive descent stack.

If less well-behaved grammars are considered, the performance results are markedly different. 
All three parsing algorithms were timed matching the grammar in Figure~\ref{exp-grammar} on strings of the form $\chs{a}^n\chs{c}^n$, with a one-minute cutoff for execution time. 
As predicted, naive recursive descent parsing had exponential runtime growth, with $n=27$ being the largest instance it could parse inside the time limit. 
Derivative parsing and packrat both showed linear growth in memory usage. 
Corresponding to this linear growth in memory usage was a quadratic growth in runtime for the derivative parsing algorithm, exceeding the time bound after $n=4000$. 
Packrat parsing was very performant on this simple example, with no instance up to $n=10000$ reporting any perceptible runtime.

\section{Conclusions and future work}
This work shows that a polynomial runtime guarantee for PEG parsing incurs significant penalties for many common inputs; the new analysis presented in Section~\ref{well-behaved-sec} provides a theoretical explanation for the good behaviour of recursive descent parsing on a restricted set of inputs, while the experimental results in Section~\ref{experimental-sec} demonstrate the real-world practicality of these restrictions on the input. 
The derivative parsing algorithm presented in this paper provides a choice of trade-offs to guarantee polynomial runtime; it may be a practical choice in memory-constrained or online systems where the asymptotic increase in memory usage incurred by packrat parsing or input buffering is undesirable.

The algorithm presented in this paper is, formally speaking, a recognition algorithm rather than a parsing algorithm, as it does not produce a parse tree. 
It should be fairly simple to augment $\varepsilon$ expressions with parse tree nodes, composing these nodes as their containing expressions matched or failed. 
An alternate approach would be to integrate semantic action expressions into the grammar; this should be possible by augmenting each parsing expression with an environment to store local variables. 
A third option would be to extend the derivative of parsing expressions defined in this paper to a derivative of parsing expression parsers, analogously to the extension defined by Might~\etal\ for CFG derivatives \cite{MDS11}. 
All these extensions are left to future work.

It would also be interesting to attempt to handle left-recursion in derivative parsing with sensible semantics. 
Ford \cite{For02} removes left-recursive rules by grammar transformation, Medeiros~\etal \cite{MMI14} modify PEG semantics to support left-recursive grammars, and Laurent and Mens \cite{LM15} use a custom parsing expression to handle left-recursion; one of these approaches could possibly be applied dynamically in the derivative parsing algorithm.

{\it The author would like to acknowledge the helpful feedback of Matt Might, David Darais, Peter Buhr, Adam Roegiest, and a number of anonymous reviewers, as well as Kota Mizushima's provision of test data and grammars. 
This work was supported in part by the author's NSERC post-graduate scholarship.}

\bibliographystyle{eptcs}
\bibliography{peg_deriv}

\end{document}